\documentclass[]{aa}

\usepackage{graphicx}

\newcommand{\kms}{\ifmmode {{\rm km\,s^{-1}}}                 
                  \else {\hbox{{\rm km$\,$s$^{\rm -1}$}}}\fi}

\begin{document}

\title{Imaging the circumstellar envelopes of AGB stars
  \thanks{Based
   on observations made at the European Southern Observatory, Chile
   (program  63.I-0177A), on de-archived data made with the NASA/ESA
   {\it Hubble Space Telescope}, and on observations made 
   at Haute-Provence Observatory, France }
 }
\author{N. Mauron\inst{1} \and  P.~J.~Huggins\inst{2}}

\offprints{N. Mauron}

\institute{Groupe d'Astrophysique, UMR 5024 CNRS, Case CC72,
Place Bataillon, F-34095 Montpellier Cedex 5, France
\email{mauron@graal.univ-montp2.fr}
\and
Physics Department, New York University, 4 Washington Place, New York
NY 10003, USA}

\date{Received xxx/  Accepted xxx}

\abstract
{}{We report the results of an exploratory program to image
the extended circumstellar envelopes of asymptotic giant branch (AGB)
stars in dust-scattered galactic light. The goal is to characterize
the morphology of the envelopes as a probe of the mass-loss process.}
{The observations consist of short exposures with the VLT and longer
exposures with 1--2~m telescopes, augmented with archival images from
the Hubble Space Telescope.}
{We observed 12 AGB stars and detected the
circumstellar envelopes in 7. The detected envelopes have mass loss
rates $\ga 5\times 10^{-6}$~$M_{\odot}$\,yr$^{-1}$, and they can be
seen out to distances $\ga 1$~kpc. The observations provide
information on the mass loss history on time scales up to $\sim
10,000$~yr.  For the five AGB envelopes in which the circumstellar
geometry is well determined by scattered light observations, all
except one (\object{OH348.2$-$19.7}) show deviations from spherical
symmetry. Two (\object{IRC+10216} and \object{IRC+10011}) show roughly spherical
envelopes at large radii but asymmetry or bipolarity close to the
star; one (\object{AFGL 2514}) shows an extended, elliptical envelope, and one
(\object{AFGL 3068}) shows a spiral pattern.  The non-spherical structures are
all consistent with the effects of binary interactions.}
{ Our observations are in accord with a scenario in which binary companions
play a role in shaping planetary nebulae, and show that the
circumstellar gas is already partly shaped on the AGB, before
evolution to the proto-planetary nebula phase.}

\keywords{ Stars: AGB and post-AGB $-$ Stars: mass loss
$-$ Stars: circumstellar matter}

\titlerunning{The circumstellar envelopes of AGB stars}
\authorrunning{N. Mauron and P. J. Huggins}
\maketitle
\section{Introduction}


\begin{table*}[!t]
\caption[]{The observed sample of AGB stars}

	\begin{center}
        \begin{tabular}{lcrrrrccrrr}
        \noalign{\smallskip}
        \hline
	\hline
        \noalign{\smallskip}

 Star & IRAS name &  $l$ & $b$  & C/O & $d$  & $V_{\rm exp}$ & $\dot{M}$               & $f_{12}$ & $f_{12}/f_{25}$ & Note\\
      &           &      &      &     & (pc) & (\kms) &(M$_{\odot}$\,yr$^{-1}$) &  (Jy)  &                 &\\                                                       
\noalign{\smallskip}
\hline
\noalign{\smallskip}
\noalign{\smallskip}

\object{IRC+10011} & 01037$+$1219& 128 & $-$50    & O &  740 & 20.7 & 2.2 10$^{-5}$  & 1155  &1.19& (1)\\
\object{YY Tri}    & 02152$+$2822 & 145 & $-$31   & C &  2100&  9.2 & 0.2 10$^{-5}$ & 121  & 1.08&\\
\object{IK Tau}    & 03507$+$1115 & 178 & $-$31   & O &  300 & 19.6 & 0.5 10$^{-5}$ & 4634 & 1.95&(1)\\
\object{IRC+70066} & 05411$+$6957 & 143 & $+$20   & O &  730 & 21.9 & 0.4 10$^{-5}$ & 801  & 1.97&(1)\\
\object{AFGL 5254} & 09116$-$2439 & 253 & $+$16   & C & 1170 & 13.1 & 1.4 10$^{-5}$ & 737  & 1.85&(1)\\
\object{CIT 6}     & 10131$+$4039 & 198 & $+$56   & C &  460 & 16.8 & 0.8 10$^{-5}$ & 3319 & 2.72&(1)\\
\object{AFGL 2155} & 18240$+$2326 &  52 & $+$15   & C & 1070 & 16.1 & 1.6 10$^{-5}$  &  731  &1.63& (1)\\
\object{OH 348.2$-$19.7} & 18467$-$4802 & 348 & $-$19   & O & 1200 & 12.3 & 0.8 10$^{-5}$  &  284  &0.83&\\
\object{AFGL 2514} & 20077$-$0625 &  36 & $-$20   & O &  660 & 16.2 & 0.8 10$^{-5}$  & 1255  &1.18&\\
\object{AFGL 3068} & 23166$+$1655 &  93 & $-$40   & C & 1080 & 14.1 & 4.2 10$^{-5}$  &  706  &0.91& (1)\\
\object{AFGL 3099} & 23257$+$1038 &  92 & $-$47   & C & 1300 & 10.1 & 0.5 10$^{-5}$  &  190  &1.34&\\
\object{AFGL 3116} & 23320$+$4316 & 108 & $-$17   & C &  870 & 14.7 & 1.1 10$^{-5}$  &  959  &2.04& (1)\\
\noalign{\smallskip}
\noalign{\smallskip}
\hline
\end{tabular}
\end{center}
{\small Notes: (1) $d$ and $\dot{M}$ from Olivier et al. (\cite{olivier01}),
  otherwise from Loup et al. (\cite{loup93})}
\end{table*}

One of the most striking aspects of the evolution of stars from the
asymptotic giant branch (AGB) to the planetary nebulae (PN) phase is
the complex structure formed in the circumstellar gas (e.g., Balick \&
Frank \cite{ballick02}).  This has been extensively observed in PNe
using high resolution optical imaging with the Hubble Space
Telescope (HST), and shows both large and small scale features such
as multiple arcs, bubbles, bicones, point-symmetric knots and bullets,
tori, ansae, and globules. Many examples can be seen in the recent
volume edited by Meixner et al.\ (\cite{meixner04}).

Some of the structural features in the nebulae are related to
photo-ionization or fast winds produced during the PN phase, but
others are formed earlier and are only partly understood. The multiple
arcs are known to result from modulation of the mass-loss during the
late AGB phase (Mauron \& Huggins \cite{mh00}), although the underlying
mechanism is unknown. Similarly, the point symmetries seen in PNe can
be traced to the effects of bipolar jets which become prominent in the
proto-PN phase (e.g., Lopez \cite{lopez03}), but the detailed picture is not
clear.  The uncertain, early development of these jets and their
possible links to other structural features such as disks or tori
which emerge from the last major episodes of mass-loss underscore the
importance of observations of the final stages of the AGB.

The detailed structure of the circumstellar envelopes of AGB stars has
not been studied in a large number of cases because the relatively
cool material is a challenge to high resolution
observations.  One technique that has been used to image the
circumstellar gas is millimeter interferometry, and the most extensive
work in this area is by Neri et al.\ (\cite{neri98}) who produced an atlas of
AGB envelopes in the molecular lines of CO. With an angular resolution
of $\sim 3\arcsec \times 5\arcsec$ and limited dynamic range, they
conclude that most (70\%) AGB envelopes are consistent with spherical
symmetry. A second technique that has been used to observe the
envelopes is imaging the thermal infrared dust emission. This is
possible at very high angular resolution, and studies of individual
objects have revealed asymmetries very close the the central stars
(e.g., Monnier et al.\ \cite{monnier04}, Weigelt et al.\ \cite{weigelt02}). 

A third technique of imaging the circumstellar dust in scattered light
at optical wavelengths has been used by us in the case of the nearby
AGB archetype IRC+10216 (Mauron \& Huggins \cite{mh99}, \cite{mh00}).  The inner
envelope is illuminated by light from the central star, and the outer
envelope is illuminated by the ambient galactic radiation field. With
ground based and HST observations, the observations provide a
resolution of 0\farcs1--1\arcsec\ over a very large field, and in the
case of IRC+10216, we were able to detect scattered light out to
200\arcsec\ from the central star and to determine the detailed
geometry of the envelope.

In this paper we report on a pilot survey to explore the extent to
which this technique can be used to study the large scale structure of
the circumstellar envelopes of much more distant AGB stars.

\section{The AGB sample}

The AGB stars observed in the survey are listed in Table 1.  They were
selected from the catalog of Loup et al.\ (\cite{loup93}) with priority given to
nearby objects with strong IRAS fluxes and large infrared excesses.
They were also chosen to lie at relatively high galactic latitudes, in
order to minimize galactic extinction and to avoid crowded background
star fields.  The sample contains 12 objects. It is not complete with
regard to any observable quantity, but should be regarded as an
exploratory sample, going beyond the single case of IRC+10216 reported
earlier.

Table~1 lists for each object the IRAS name, the galactic
co-ordinates, the chemistry (oxygen-rich or carbon-rich), distance
$d$, expansion velocity $V_{\rm exp}$, mass-loss rate $\dot{M}$, and
relevant IRAS fluxes from the point source catalog. If available, the
values of $d$ and $\dot{M}$ are taken from Olivier et al.\ (\cite{olivier01}) in
which distances were determined through the period luminosity
relation, otherwise they are from Loup et al.\ (\cite{loup93}).

\section{Observations}

Details of the observations including the telescopes, the filters,
and the exposure times are listed in Table~2.

\object{IRC+10011}, \object{OH 348.2$-$19.7}, and \object{AFGL 5254}
 were observed with the ESO
VLT using the FORS1 focal reducer. Relatively short exposures were
made in U, B, \& V filters.  The images are ${6\farcm8} \times
6\farcm8$ with a pixel size of $0\farcs200$, and the image quality is
generally near $1\farcs1$ (FWHM).

\object{AFGL~2155}, \object{AFGL~3068}, \object{AFGL~3099}, 
\object{AFGL~3116}, \object{YY~Tri}, \object{IK~Tau}, \object{IRC+70066}
and \object{CIT6} were observed with the 1.20~m telescope of the Observatoire de
Haute-Provence (OHP). This telescope is equipped with a camera which
provides a field of $11\farcm7 \times 11\farcm7$ with a pixel size of
$0\farcs684$.  In cases where the object was not easily identified, the
observations were started with a short I band exposure which typically
results in a clear identification, followed by longer exposures in B
and V. In a few cases U band exposures were also obtained.  The image
quality is $\sim 2$--3$\arcsec$ (FWHM).

\object{AFGL 2514} was observed with the ESO Danish 1.54~m telescope using the
DFOSC focal reducer, and only in the V band. The image size is
$13\farcm7 \times 13\farcm{7}$ with a pixel size $0\farcs403$. The
image quality is {1\farcs0}.

All CCD frames were reduced using standard procedures including bias
subtraction, flat-field correction, and cosmic-ray rejection.  When
several frames were made with the same passband, the frames
were appropriately shifted and summed.

Where available we also made use of images from the HST archives to
complement our observations at high resolution. Relevant data was
found in four cases, and details are given in Table~3. These data
consist of ACS images, with a field size of $211\arcsec \times
211\arcsec$ and scale {0\farcs050} per pixel. Data reduction included
co-addition of the images, correction for cosmic ray hits, and rotation for
direct comparison with the ground based images. As far as we are aware,
none of these HST data have previously been published.

\begin{table*}[!t]
\caption[]{Details of the Observations}

	\begin{center}
        \begin{tabular}{lrlllll}
        \noalign{\smallskip}
        \hline
	\hline
        \noalign{\smallskip}

 Star & Telescope & Band \& Exposure & V/B Image$^a$ &  Surface Brightness     & Notes\\
      &           &     (min)                   & &  (V-mag arcsec$^{-2}$ ) &\\                                                   
\noalign{\smallskip}
\hline
\noalign{\smallskip}
\noalign{\smallskip}
\object{IRC+10011} &  VLT\, 8.00 m & U (4), B (2), V (2)      &
      st + env   & 25.7 (at  5\arcsec)      &  1 \\
\object{YY Tri}    & OHP 1.20 m    & B (240), V (120) & $\ldots$.  & $<$ 25.1      &  \\
\object{IK Tau}    & OHP 1.20 m    & U (90), B(150)           & st  & $<$
      24.9 (at 10\arcsec)  &  2  \\ 
\object{IRC+70066} & OHP 1.20 m    & U (480), B (420), V (330)& st  & $<$ 25.8 (at 10\arcsec)  &  \\
\object{AFGL 5254} & VLT\, 8.00 m  & U (4), B (2), V (2)      & st  & $<$
      25.4 (at 5\arcsec)   &  \\
\object{CIT 6}     & OHP 1.20 m    & U (102), B (120)         & st  & $<$
       24.9 (at 10\arcsec)  &  2 \\
\object{AFGL 2155} &  OHP 1.20 m   & V (300)          & env & 25.3          &  \\
\object{OH\,348.2$-$19.7} &  VLT\, 8.00 m & U (8), B (2), V (2)      &
      env  & 24.9
      (at 5 \arcsec)      &  1, 3 \\
\object{AFGL 2514} &  ESO\, 1.54 m & V (120)                 & env  & 23.6
         &  1 \\ 
\object{AFGL 3068} &  OHP 1.20 m   & B (60), V (60)          & env & 23.9
             & 1 \\
\object{AFGL 3099} &  OHP 1.20 m   & V (240)                 & env & 26.0
         &  \\
\object{AFGL 3116} &  OHP 1.20 m   & B (150), V (60)         & st + env
      & 25.8 &  2, 4 \\
\noalign{\smallskip}
\noalign{\smallskip}

\noalign{\smallskip}
\hline
\end{tabular}
\end{center}
{\small $^a$ Detections in the V and B images: st\,=\,star,
  env\,=\,envelope. The star \object{YY Tri} is seen only in the I-band. \\ 
Notes: (1) See Table~3 for HST data. (2) Surface brightness
  from B image, assuming B$-$V = 0.5.
(3) Star near center is field star. (4) Envelope seen only in B image.}
\end{table*}

\begin{table}[!t]
\caption[]{Details of the HST Archival Observations}

	\begin{center}
        \begin{tabular}{llll}
        \noalign{\smallskip}
        \hline
	\hline
        \noalign{\smallskip}

 Star & Dataset   & Filter    &  Exposure\\
      &           &      &  (min)  \\                                                   
\noalign{\smallskip}
\hline
\noalign{\smallskip}
\noalign{\smallskip}
\object{IRC+10011} & j8di89dmq   & F814W  & 11.3 \\
                   & j8di89drq   &        &      \\
\object{OH\,348.2$-$19.7} & j92k41doq   & F606W  & 11.6 \\
                   & j92k41dtq   &        &      \\ 
\object{AFGL 2514} & j92k52i4q   & F606W  & 11.6 \\ 
                   & j92k52i9q   &        &      \\
\object{AFGL 3068} & j92k58h0q   & F606W  & 23.1 \\
                   & j92k58h1q   &        &      \\
                   & j92k58h2q   &        &      \\
                   & j92k58h3q   &        &      \\
\noalign{\smallskip}
\noalign{\smallskip}

\noalign{\smallskip}
\hline
\end{tabular}
\end{center}
\end{table}


\section{The circumstellar envelopes}

\subsection{The images}
The circumstellar envelopes were detected in scattered light around 7
of the AGB stars observed, and not detected around 5. Images and
intensity profiles are discussed individually in \S 4.3.

The images shows that the envelopes can be illuminated as in \object{IRC+10216}
in two ways, by the central star or by the ambient galactic radiation
field. In the first case, the central star forms a small scattering
core whose intensity decreases rapidly with angular distance from the
center ($\sim \theta^{-3}$, Martin \& Rogers \cite{martin87}. In the second
case, a faint, extended nebula is seen, with a relatively shallow
radial dependence (see below), and it may form a plateau in the
central regions where the optical depth to external radiation exceeds
$\sim 1$.  Both effects may occur at the same time. The typical
situation in the thicker envelopes is that the star and the
star-illuminated core are faint or not seen in the B or V bands, but
the star becomes dominant at longer wavelengths because of its low
temperature and the decreasing opacity of the dust envelope.

   \begin{figure*}[!ht]
    \resizebox{\hsize}{!}{
    {\rotatebox{0}{\includegraphics{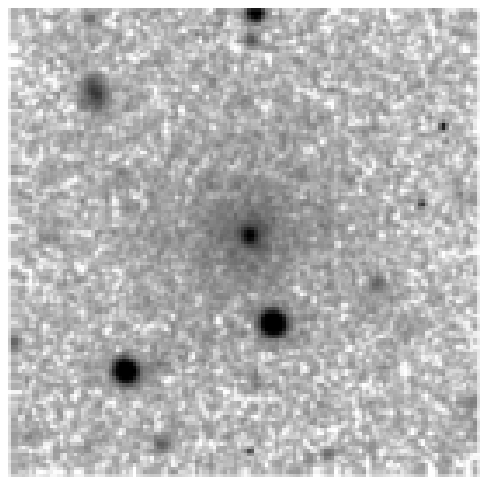}}}\hspace{0.3cm}
    {\rotatebox{0}{\includegraphics{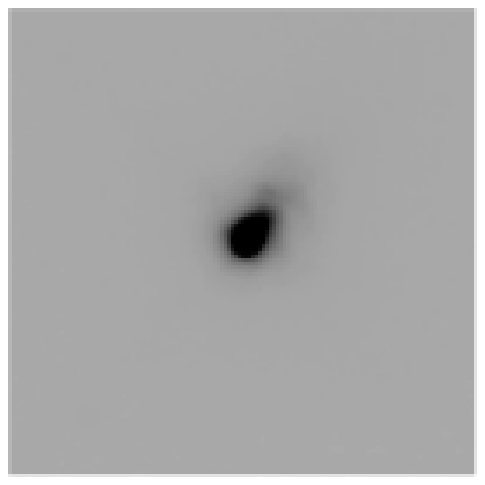}}} }
    
    \caption[]{\object{IRC+10011}. \emph{Left:} VLT image in the V-band. 
     Field size: $51\farcs2 \times 51\farcs2$.
     \emph{Right:} HST-ACS image in the F816W filter. 
     Field size: $6\farcs4 \times 6\farcs4$.
     North is up and east is to the left in all images.
      }
     \label{fig01}
     \end{figure*}


 \begin{figure*}[!ht]   
  \centerline{
     \includegraphics[origin=c,angle=-90,width=0.48\linewidth]{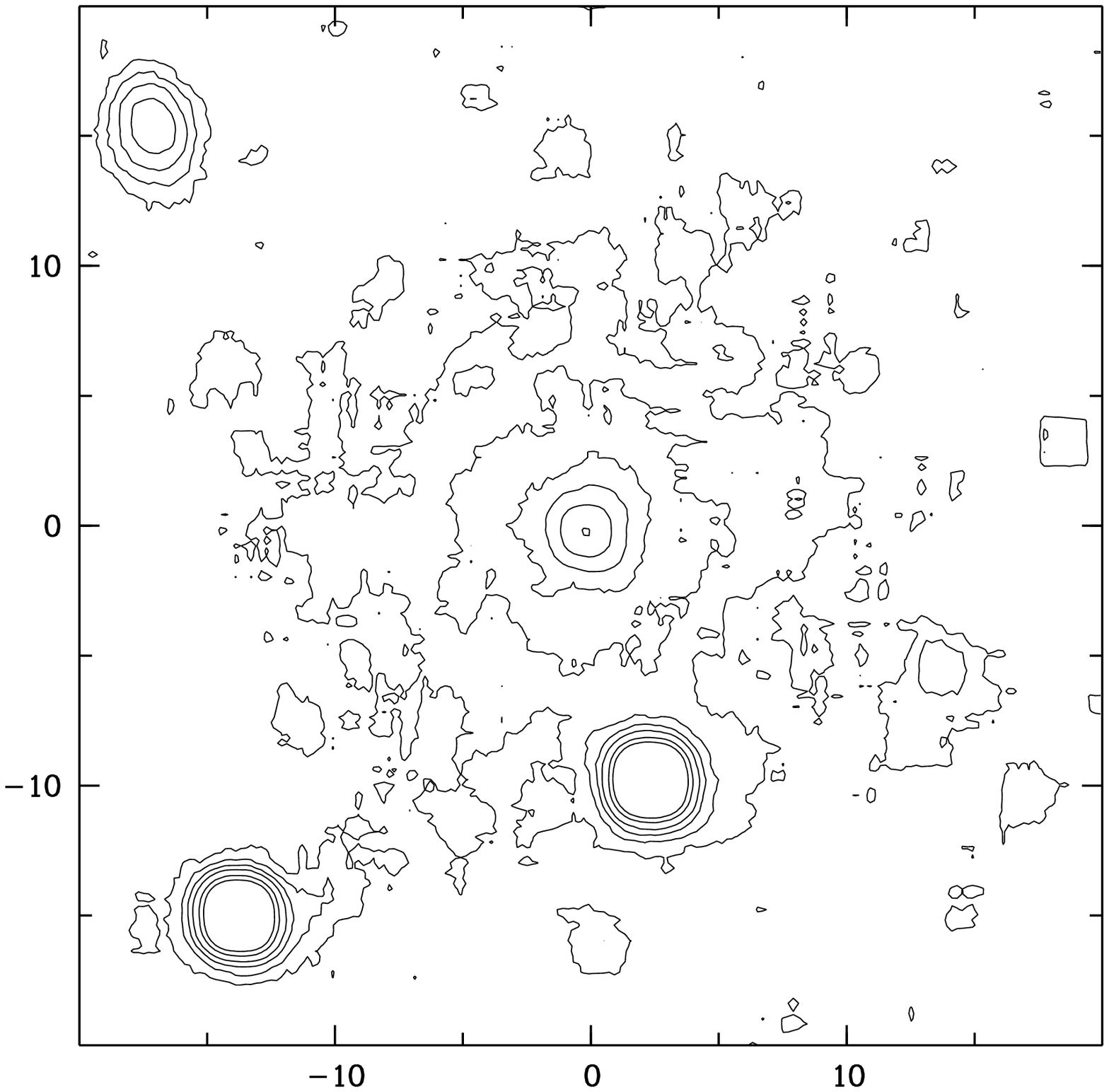}\hspace{0.4cm}
    \includegraphics[width=.48\linewidth]{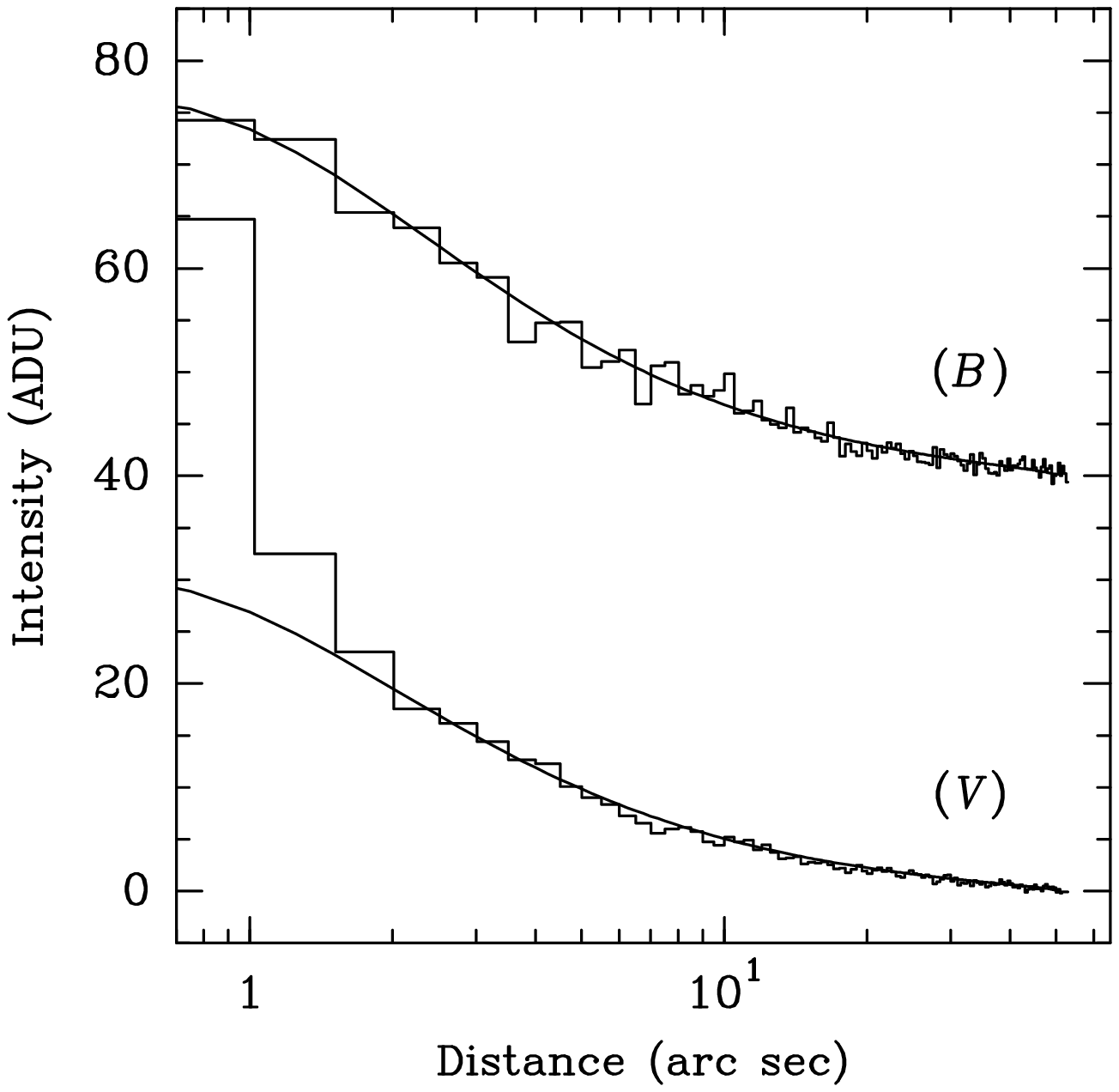} }
    
    \caption[]{ \object{IRC+10011}. \emph{Left}: Close-up isophotes of the VLT
     V-band image. Field size: $40\arcsec \times 40\arcsec$.
       \emph{
     Right}: Radial profiles, averaged in azimuth, from the VLT images
     in B and V. In the V profile, 1 ADU above the background corresponds 
     to 2.0 $\times $ 10$^{-16}$ erg\,s$^{-1}$\,cm$^{-2}$\,$\mu$$^{-1}$\,arcsec$^{-2}$,
     or 28.2 V-mag\,arcsec$^{-2}$. No calibration was possible for the B profile.
      The B data are scaled by a factor of 4.2 and offset
     from zero for clarity. The solid lines show model fits to the
     profiles for a constant mass loss rate. See text for details.}
     
  \label{fig02}   
  \end{figure*}

\subsection{Surface brightness}

For each circumstellar envelope we determined the V-band surface
brightness, either at the center if no core was
visible, or at an offset away from the center if the core was
seen. The results are given in column 5 of Table~2. Upper limits are
given for the cases where the extended envelope was not detected. For
comparison, the  V-band surface brightness of the central
plateau of \object{IRC+10216} is 25.2 mag arcsec$^{-2}$ (Mauron et al. \cite{mauron03}).

Where possible, the surface brightness of the envelopes was calibrated
by using counts of photometric standard stars imaged on the same
night.  Where this was not possible, the image calibration was based
on photometric data for field stars from the APM catalog (Irwin \cite{irwin00}).
By considering the blue and red APM photometry for several sequences
with Johnson UBV photometry, it was possible to derive the following
relations between Johnson and APM magnitudes:\begin{displaymath} V
\approx 0.5*(B_{APM}+R_{APM}) -0.4,\;\mathrm{and} \end{displaymath}
\begin{displaymath}
B \approx B_{APM}-0.2.\end{displaymath} with a scatter for these
relations of $\sim$ 0.2 mag.  There is good consistency in cases
where both methods can be applied to the same envelope, but because
the envelopes are faint, i.e., a few percent of the sky brightness,
the quality of the surface brightness measurements is also affected by
sky noise, and the final error bars are estimated to be around
$\pm$0.4 mag. In the case of \object{IK~Tau}, \object{CIT~6}, and 
\object{AFGL~3116}, the
V-band brightness was estimated from the B-band image, using
the assumption that  B$-$V= 0.5, which is the color measured for
\object{IRC+10216} (Mauron et al.\ \cite{mauron03}).

\subsection{Individual envelopes}

\subsubsection{ \object{IRC+10011} (\object{WX Psc}, \object{CIT~3})}

The V-band image of \object{IRC+10011} obtained with the VLT is shown in the
left hand panel of Fig.\,\ref{fig01}. It shows a bright core illuminated by the
central star, and a faint, extended envelope illuminated by the
ambient radiation field. The U and B-band images are similar, although
the central core is suppressed at the shorter wavelengths.

The images of the extended envelope appear approximately circularly
symmetric, and this is seen more clearly in the close-up isophotal
plot of the V-band image in Fig.\,\ref{fig02} (left). Intensity strips in azimuth,
averaged in sectors of 20\degr\ show variations of less than 15\%. The
radial variation of the envelope in B and V (averaged in azimuth) is
shown in Fig.\,\ref{fig02} (right).  The envelope can be traced out to $\sim 50\arcsec$;
the limitation is set by variations in the background level.

The HST-ACS image of \object{IRC+10011} in the far red F818W filter is shown in
the right hand panel of Fig.\,\ref{fig01}.  In this image the star is dominant,
and only the innermost regions of the envelope are seen. To the NW at
$\sim 0\farcs8$ from the center there is part of a faint circular arc,
probably similar to the arcs seen in \object{IRC+10216}. However, in contrast
to the outer envelope, the core structure is highly asymmetric, with a
bright extension out to $\sim {0\farcs4}$ at $PA \sim -45\degr$.

The inner envelope of \object{IRC+10011} has also been studied at high
resolution by Hofmann et al.\ (\cite{hofmann01}) using bispectrum speckle
interferometry in the J, H, and K bands.  The core is found to be
point-like in H and K, but in J it is elongated along a
symmetry axis at $PA \sim -28$\degr\ out to distances of $\sim
200$~mas. It seems most likely that the HST image captures the
extension of this core asymmetry.  Thus \object{IRC+10011} is somewhat similar
to the case of \object{IRC+10216} where approximate circular symmetry and
shells in the extended envelope co-exist with a strong axial symmetry
close to the star.


   \begin{figure*}[!ht]
    \resizebox{\hsize}{!}{
    {\rotatebox{-00}{\includegraphics{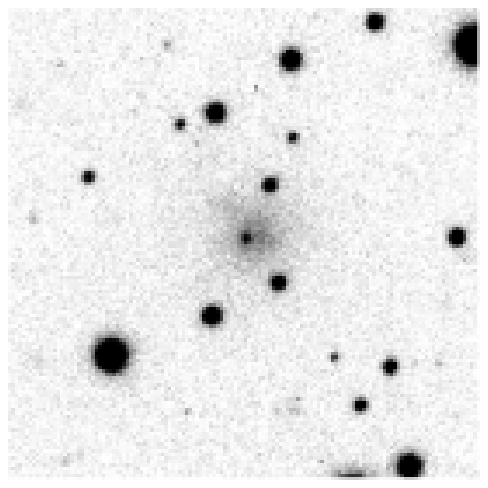}}}\hspace{0.3cm}
    {\rotatebox{-00}{\includegraphics{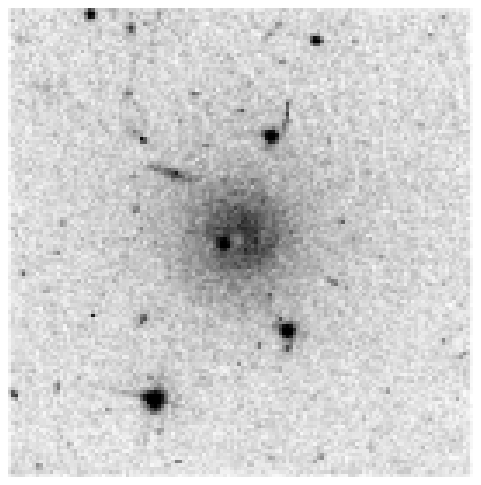}}} }
    
    \caption[]{\object{OH 348.2$-$19.7}. \emph{Left}: VLT image in the V-band. 
      Field size: $51\farcs2 \times 51\farcs2$.
     \emph{Right}: HST-ACS image in the F606W filter.
      Field size: $ {25\farcs6}\arcsec \times {25\farcs6} $.
      }
     \label{fig03}
     \end{figure*}

 \begin{figure*}[!ht]
 \centerline{
   \includegraphics[origin=c,angle=-90,width=0.48\linewidth]{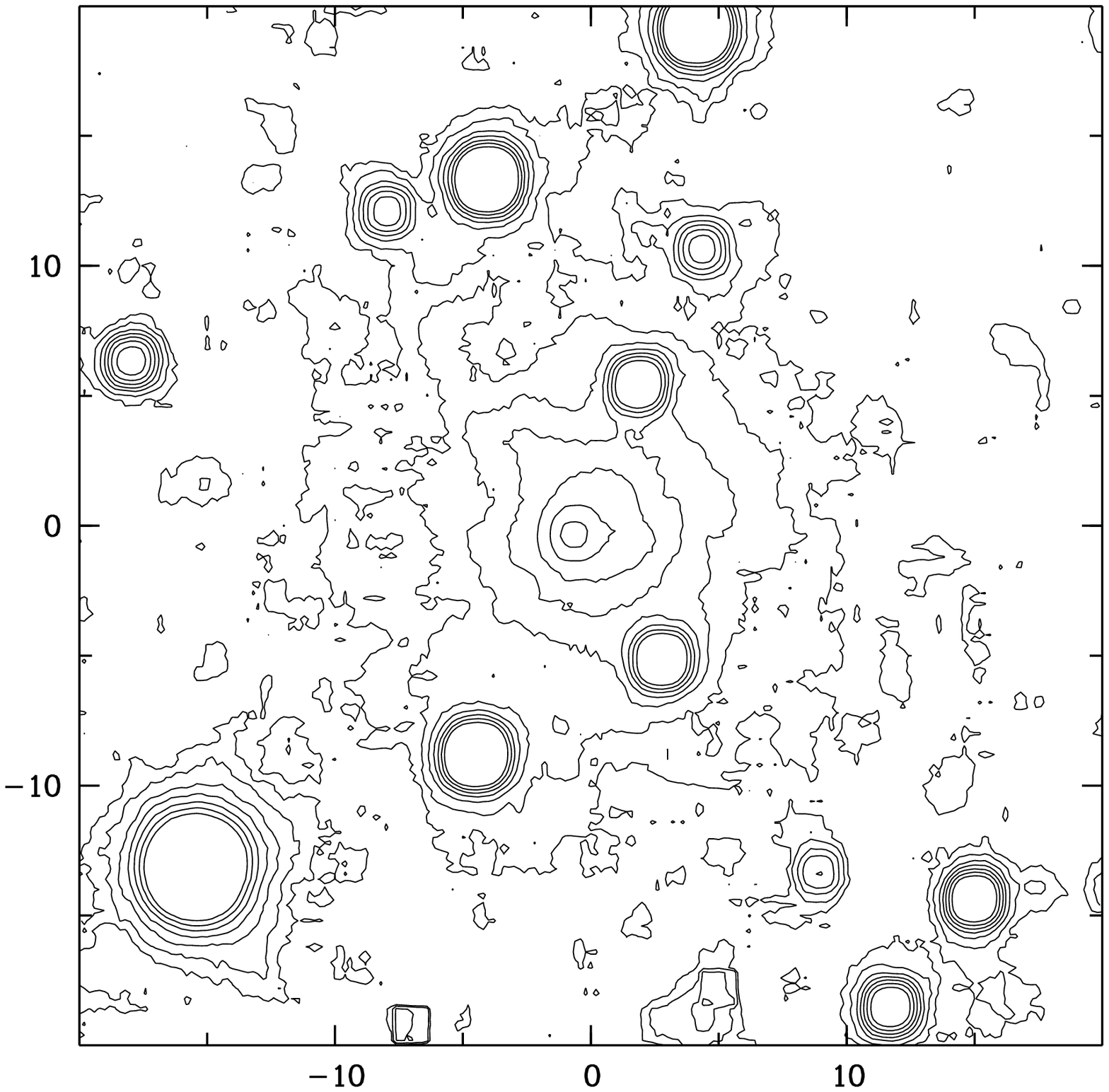}\hspace{0.4cm}
   \includegraphics[width=0.48\linewidth]{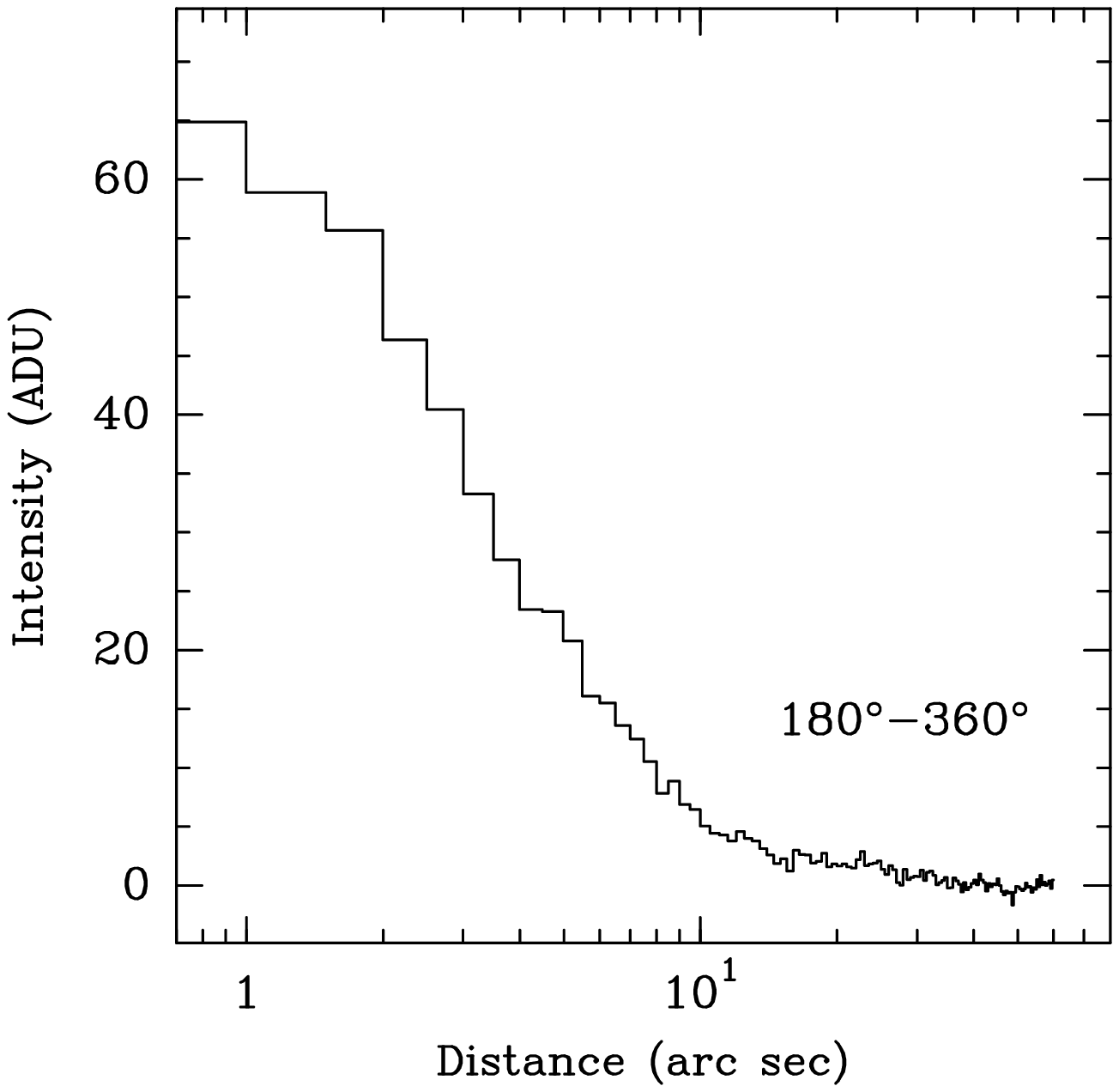}
   }
 \caption[]{\object{OH 348.2$-$19.7}. \emph{Left}: Close-up isophotes of the
 VLT V-band image. Field size: $40\arcsec \times 40\arcsec$.
 \emph{Right}: Radial intensity profile from the V-band image. In this  profile,
 1\,ADU above the background corresponds to 2.0 $\times$ 10$^{-16}$ 
   erg\,s$^{-1}$\,cm$^{-2}$\,$\mu$$^{-1}$\,arcsec$^{-2}$, or 28.2 V-mag\,arcsec$^{-2}$.} 
   \label{fig04}
 \end{figure*}

\subsubsection{\object{OH 348.2$-$19.7}}

The V-band image of \object{OH 348.2$-$19.7} obtained with the VLT is shown in
the left hand panel of Fig.\,\ref{fig03}.  The B-band image is similar. 
The extended envelope is clearly detected in both images and at first
sight looks unusual because it is not centered on the star within the
nebula, and there is an extension to the north-east.

The HST-ACS red (F606W) image is shown in the right hand panel of
Fig.\,\ref{fig03}. It too shows a star offset from the center of the nebula, 
but reveals that the extension to the north-east in the VLT image is a
background galaxy.  The issue of the star is resolved at longer
wavelengths where 2MASS images and a far red (F814W) HST image show
that the true central star is not seen in Fig.\,\ref{fig03}. It lies $\sim
0\farcs65$ to the west of the visible star, which is a foreground or
background object. This conclusion is supported by the absence of core
brightening in the nebula around the visible star.

When allowance is made for these features of the VLT images of \object{OH
348.2$-$19.7}, the overall appearance of the extended envelope is
approximately circular, as seen in the close-up isophotal image in
Fig.\,\ref{fig04} (left). To the west where the field is clearest, the envelope
can be traced out to distances of $28\arcsec$ (Fig.\,\ref{fig04}, right).  The
HST image shows that approximate circular symmetry continues into the
inner regions. It shows a dip in the intensity at the center, which is
expected for an optically thick, externally illuminated envelope, as
is seen in the case of \object{IRC+10216} (Mauron \& Huggins \cite{mh99}). 
There is also a slight east-west asymmetry in the brightness which suggests a
slightly asymmetric illumination. 

  \begin{figure*}[!ht]
    \resizebox{\hsize}{!}{
    {\rotatebox{-00}{\includegraphics{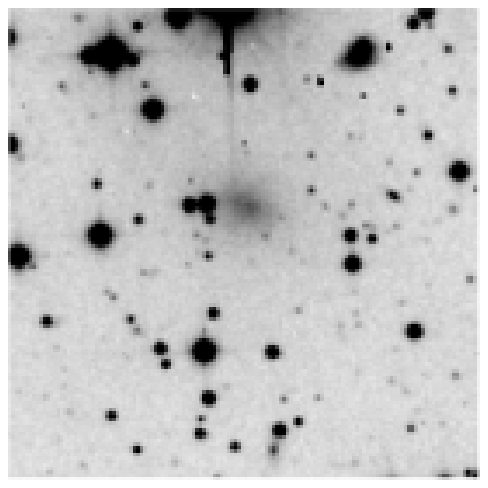}}}\hspace{0.3cm}
    {\rotatebox{-00}{\includegraphics{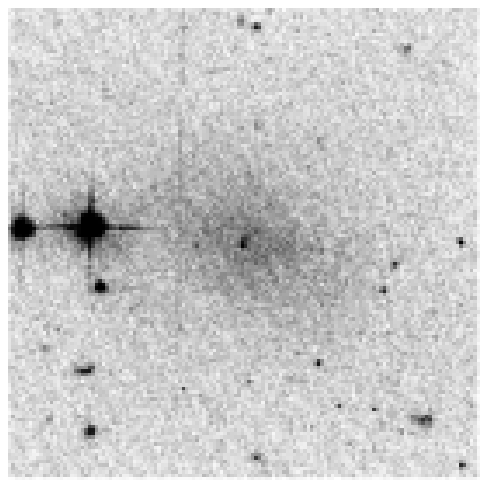}}} }
     \caption[]{ \object{AFGL 2514}. \emph{Left}: ESO 1.54~m
       image in the V-band.  
     Field size: $102\arcsec \times$ 102\arcsec. 
     \emph{Right}: HST-ACS image in the F606W filter. 
     Field size: $25\farcs6 \times 25\farcs$6.
     }
     \label{fig05}
     \end{figure*}

 \begin{figure*}[!ht]
 \centerline{
   \includegraphics[origin=c,angle=-90,width=0.48\linewidth]{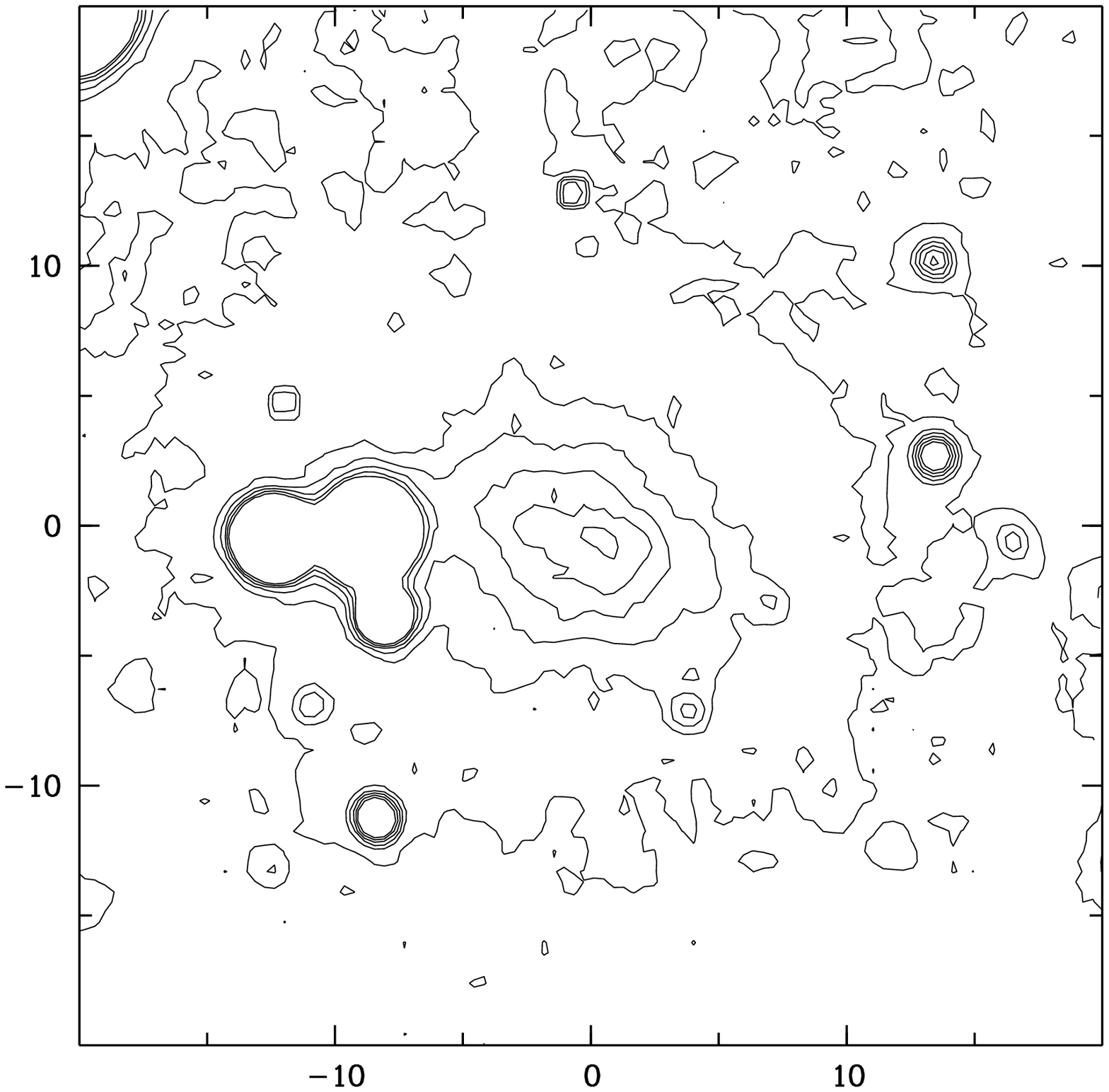}\hspace{0.4cm}
   \includegraphics[width=0.48\linewidth]{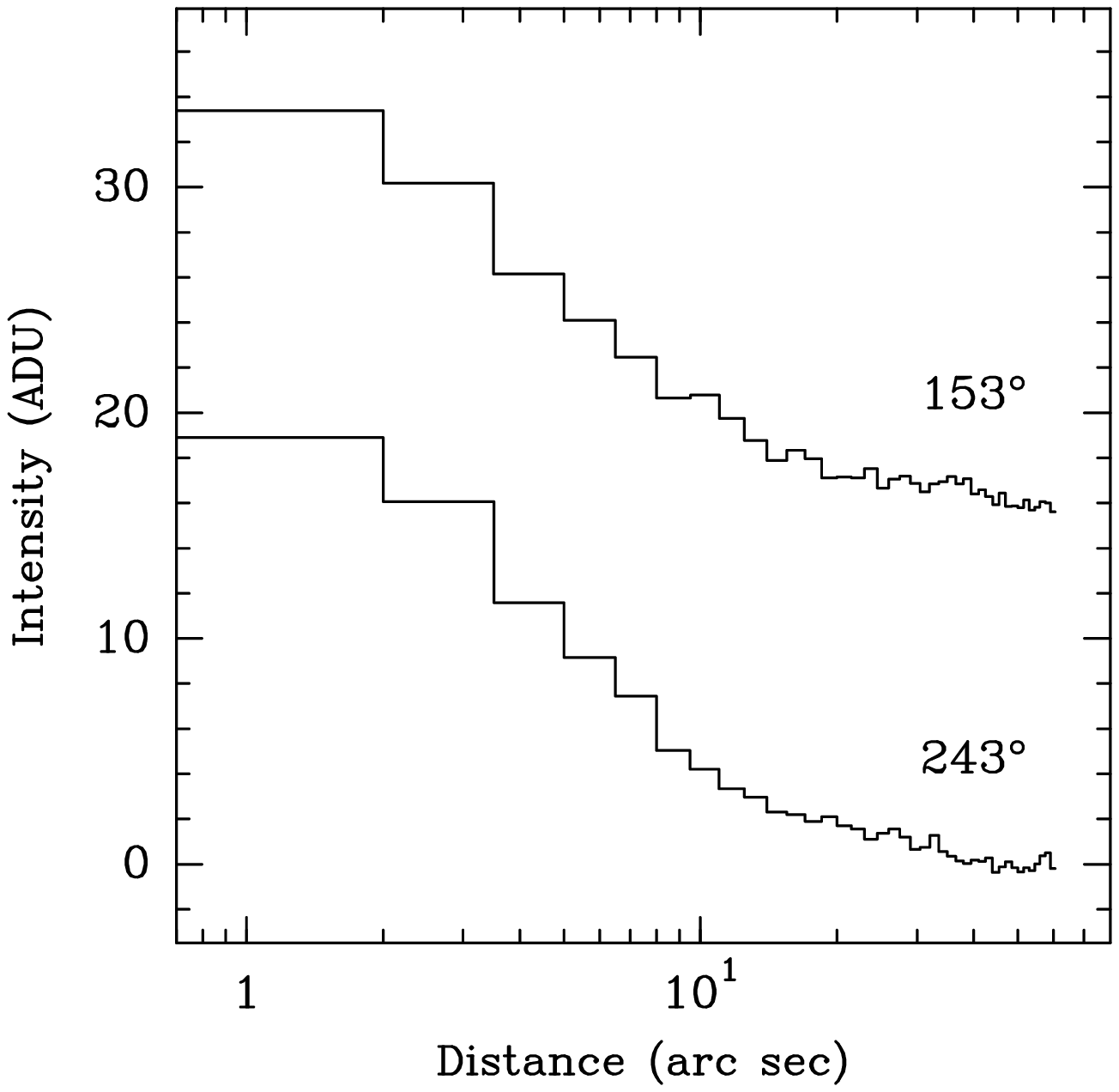}
   }
 \caption[]{ \object{AFGL 2514}. \emph{Left}: Close-up isophotes of the ESO
 1.54~m V-band image. Field size: $40\arcsec\times 40\arcsec$
 \emph{Right}: Radial profiles in the V-band along the major and
 minor axes. The intensities are scaled 
 down by 100, and the upper profile is offset for clarity. The top of both
 profiles corresponds to 1.4 $\times$ 10$^{-14}$\,erg\,s$^{-1}$\,cm$^{-2}$\,$\mu$$^{-1}
 $\,arcsec$^{-2}$, or 23.6 mag-V\,arcsec$^{-2}$}
 
 \label{fig06}
 \end{figure*}

\subsubsection{\object{AFGL 2514}}

The V-band image of \object{AFGL 2514} obtained with the ESO 1.54~m telescope
is shown in the left hand panel of Fig.\,\ref{fig05}. This is the brightest
envelope detected in the survey (Table~2), even though it is not
exceptional with regard to color or mass loss rate relative to the
other objects. No central core is seen, but the envelope is detected
with a good signal-to-noise ratio.

The envelope shows a remarkably non-spherical morphology, which is
seen more clearly in the close-up isophotal image in Fig.\,\ref{fig06} (left). The
contours are approximately elliptical, with the ratio of the major to
minor axis of $\sim 1.4$ There is no evidence to suggest that the
shape is due to asymmetric illumination. Radial intensity strips
(Fig.\,\ref{fig06}, right) show that the envelope can be detected out to 
$\sim 40\arcsec$.

The HST image of \object{AFGL 2514} in the red F606W filter is shown in the
right hand panel of Fig.\,\ref{fig05}.  It is affected by pattern noise, but is
of sufficient quality to confirm the elliptical shape of the envelope.
This image shows a point-like object at the center of the nebula,
and a further HST image in the far red F814W filter shows it to be
brighter and slightly extended. This is confirmed to be the AGB star
by comparison with 2MASS images.

  \begin{figure*}[!ht]
   
    \resizebox{\hsize}{!}{
    {\rotatebox{-00}{\includegraphics{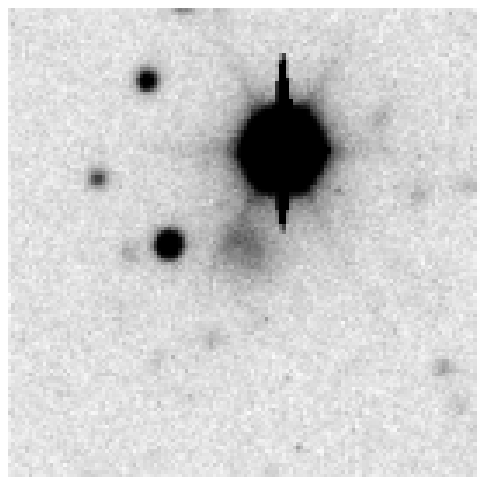}}}\hspace{0.3cm}
    {\rotatebox{-00}{\includegraphics{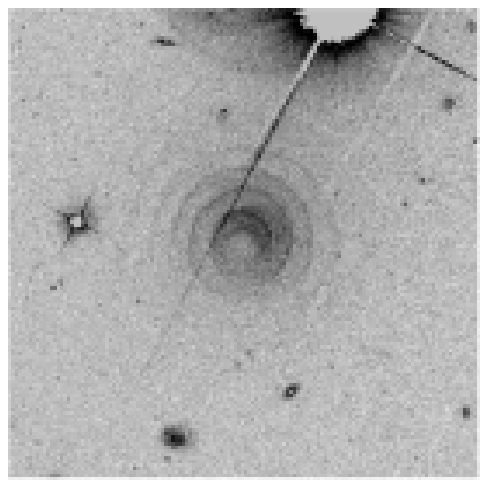}}} }
     \caption[]{ \object{AFGL 3068}. \emph{Left}: OHP 1.20~m image in the V-band.
      Field size: $117\arcsec \times 117\arcsec$.           
      \emph{Right}: HST-ACS image in the F606W filter. 
      Field size: $51\farcs2 \times 51\farcs2$.}
      
     \label{fig07}
     
\end{figure*}

   \begin{figure*}[!ht]
   
    \resizebox{\hsize}{!}{
    {\includegraphics[origin=c,angle=-00,width=0.49\linewidth]{4739f08a.ps}}\hspace{0.3cm} 
    {\includegraphics[origin=c,angle=-00,width=0.49\linewidth]{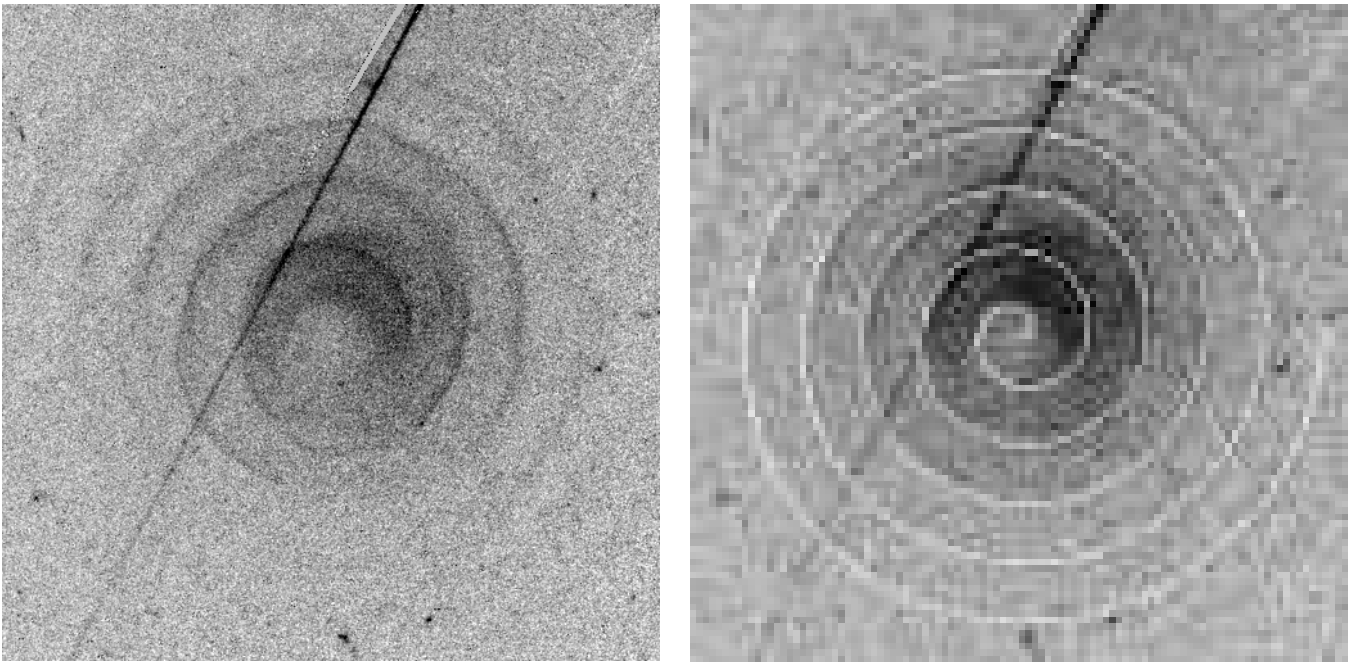}}}
     \caption[]{\object{AFGL 3068}. \emph{Left}:  Close-up of HST-ACS image
     in the F606W filter. Field size: $25\farcs6 \times 25\farcs6$. 
     \emph{Right}: Same image, with Archimedes spiral fit. See text
     for details. {\it This Fig. is available in color in the on-line version of the
     paper}
     }
     \label{fig08}
     
    \end{figure*}

  \begin{figure}[!ht]
     \resizebox{7.5cm}{!}{\rotatebox{0}{\includegraphics{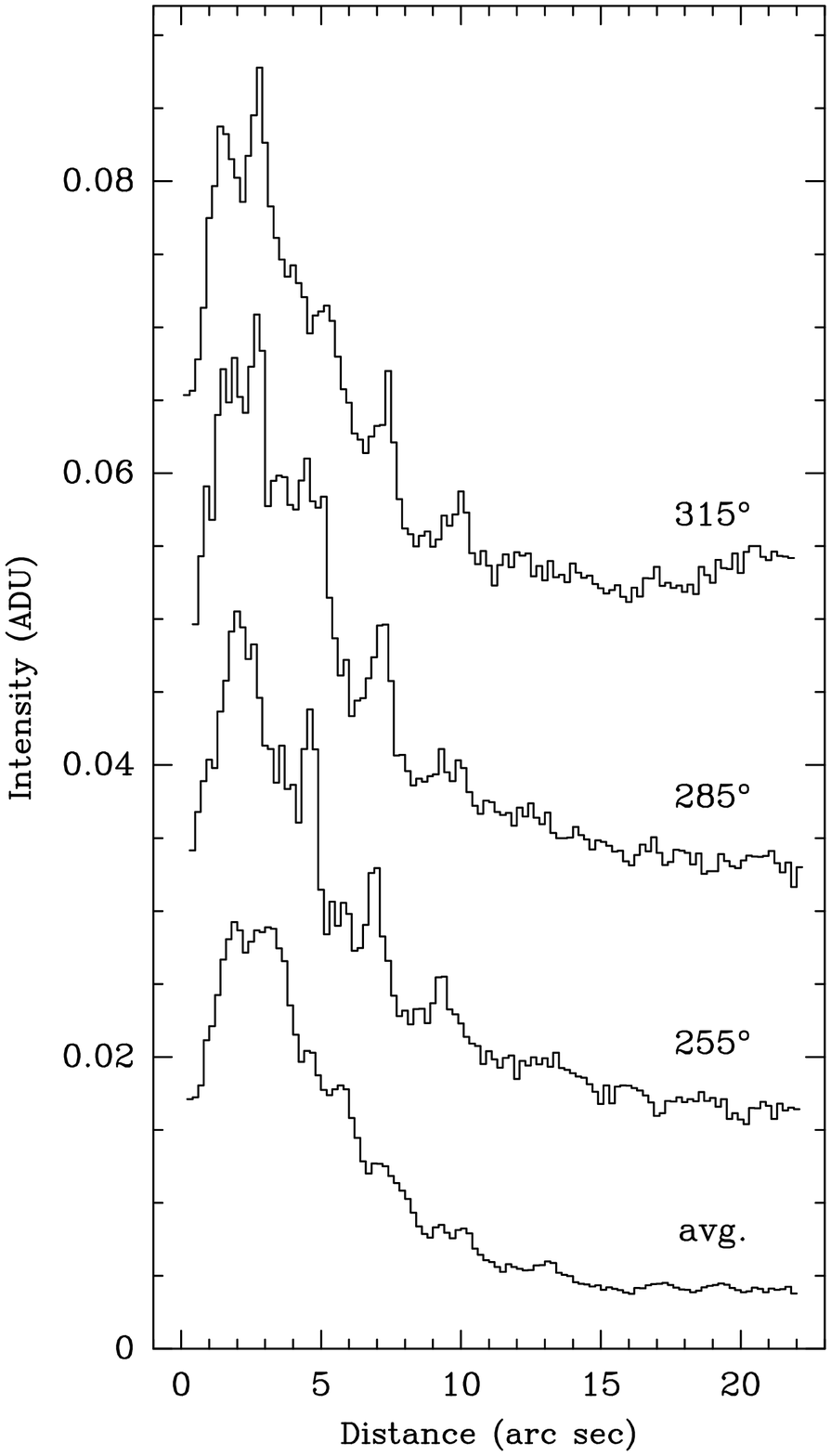}}}   
    \caption[]{\object{AFGL 3068}. Radial profiles in the F606W filter. The top
      three profiles are for adjacent 30\degr\ sectors and are
      labeled with the position angle. The maximum of the bottom curve corresponds
      to 1.1 $\times$ 10$^{-14}$ erg\,s$^{-1}$\,cm$^{-2}$\,$\mu$$^{-1}$\,arcsec$^{-2}$, or
      23.9 V-mag\,arcsec$^{-2}$.}
      
     \label{fig09}
   \end{figure}

\subsubsection{\object{AFGL 3068}}

The V-band image of \object{AFGL~3068} obtained with the OHP 1.20~m telescope
is shown in Fig.\,\ref{fig07} (left). The B-band image is similar. Neither shows
the central star, but the envelope is well detected in both
images. The scattered light can be detected out to a distance of $\sim
40\arcsec$ from the center, and the distribution is
asymmetric. Although the diffraction spike of the bright star in the
field covers part of the scattered light image, the envelope is seen
to be convex to the north-west at $PA \sim -30\degr$, suggesting that
a component of the external illumination comes from this direction.

The HST image of \object{AFGL~3068} in the red F606W filter (Fig.\,\ref{fig07}, right)
confirms the directed external illumination, and in addition reveals a
remarkable spiral pattern in the envelope (see also Appendix~A). No
central star is seen, but this becomes visible at longer wavelengths
in 2MASS images and in an HST far red F814W image. The star lies at
the center of the pattern.

The spiral pattern is single-armed, and can be traced out to a
distance of 12\arcsec\ from the center (Fig.\,\ref{fig08}; {\it this Fig.
is available in color in the on-line version of the paper}). Radial intensity profiles in
consecutive azimuthal sectors are shown in Fig.\,\ref{fig09}.  The widths of the
spiral feature are partly resolved and the peak intensities exceed the adjacent
lines of sight by factors of up to $\sim 2.5$. It can be seen in
Fig.\,\ref{fig09} that the radial profiles of the spiral are typically steep on the
side away from the star and less steep on the side toward the star,
which is characteristic of thin shells, as described by Mauron \&
Huggins (\cite{mh00}). Using a simple model of nested shells to interpret the
profiles, we find that the widths are $\la {0\farcs5}$ and the
amplitudes correspond to density contrasts between the
arm and the inter-arm material of factors up to $\sim 5$.

The pattern itself closely resembles a true Archimedes spiral, in
which the inter-arm spacing is constant. The equation for this is
$\theta_{\rm} = A\phi +C$, where $\theta$ is the angular distance from
the center, $\phi$ is the azimuthal angle (measured counter clockwise
from the west), and $A$ and $C$ are
constants which determine the inter-arm separation and the
orientation on the sky, respectively.  A least squares fit to the
pattern gives:
\[ \theta_{\rm} = 0.364(\pm0.001)\phi + 0.38 (\pm 0.01) \] 
where $\theta$ is in arcseconds, and $\phi$ is in radians. The fit is
shown superposed on the image in Fig.\,\ref{fig08} (right), where the origin of the
spiral corresponds to the location of the central star. 
The generation of the spiral is discussed in \S5.3.

\subsubsection{\object{AFGL 2155}, \object{AFGL 3099}, and \object{AFGL 3116}}

\object{AFGL~2155}, \object{AFGL~3099}, and \object{AFGL~3116}
 are three carbon-rich AGB stars
which were observed only with the OHP 1.20~m telescope.  They were
first observed in the I-band which showed bright point-like cores
allowing identification by comparison with digitized red POSS
images. Exposures in B and/or V were then made in order to image the
envelope.

\begin{figure*}[!ht]
   \centerline{
   \includegraphics[width=0.48\linewidth]{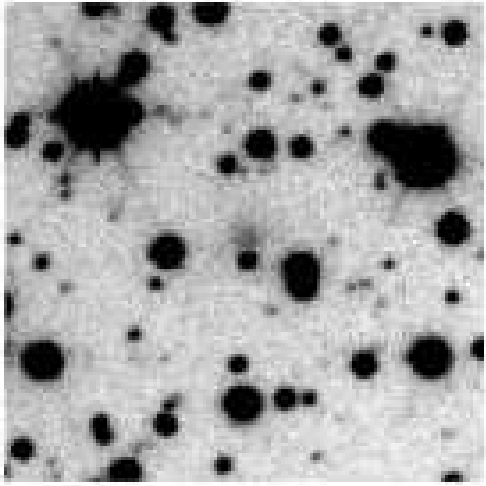}\hspace{0.2cm}
   \includegraphics[origin=c,angle=-90,width=0.48\linewidth]{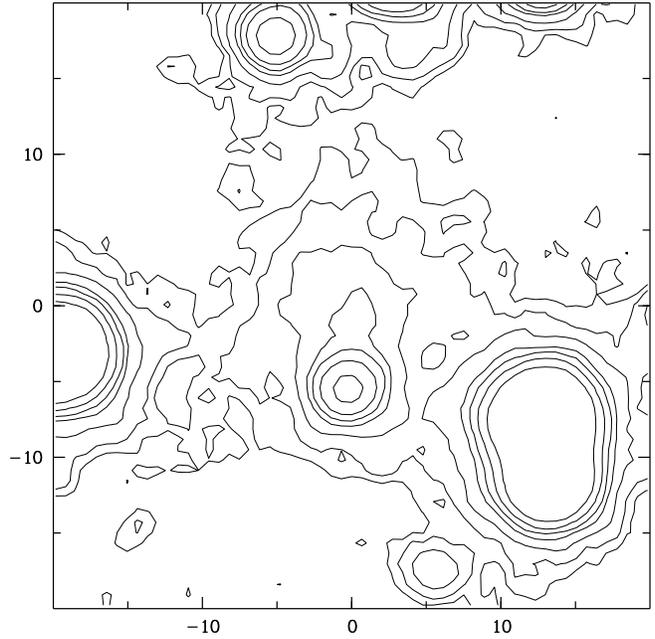} }
   
        \caption[]{ \object{AFGL 2155}. \emph{Left}: OHP 1.20~m image in the V-band. 
      Field size is $117\arcsec \times 117\arcsec$.
      \emph{Right}: Close-up isophotes of the V-band image. Field size
      is $40\arcsec \times 40\arcsec$. 
      }
     \label{fig10}
     \end{figure*}

\object{AFGL~2155} (Fig.\,\ref{fig10}) shows a bright envelope which is well detected in
the V band image. It shows no stellar core, and the envelope can be
traced out to a radial distance of $\sim 18\arcsec$ from the center.
However, on account of its low galactic latitude, it lies in a
relatively crowded star field and a foreground or background star near
the center masks the large scale geometry of the envelope at low
intensity levels.

\begin{figure*}[!ht]
   
   \centerline{
   \includegraphics[width=0.48\linewidth]{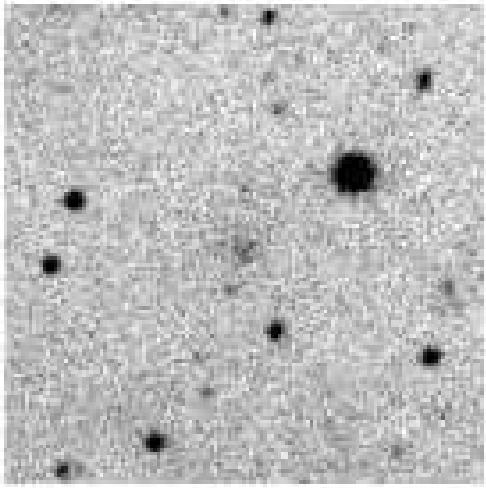}\hspace{0.2cm}
   \includegraphics[origin=c,angle=-90,width=0.48\linewidth]{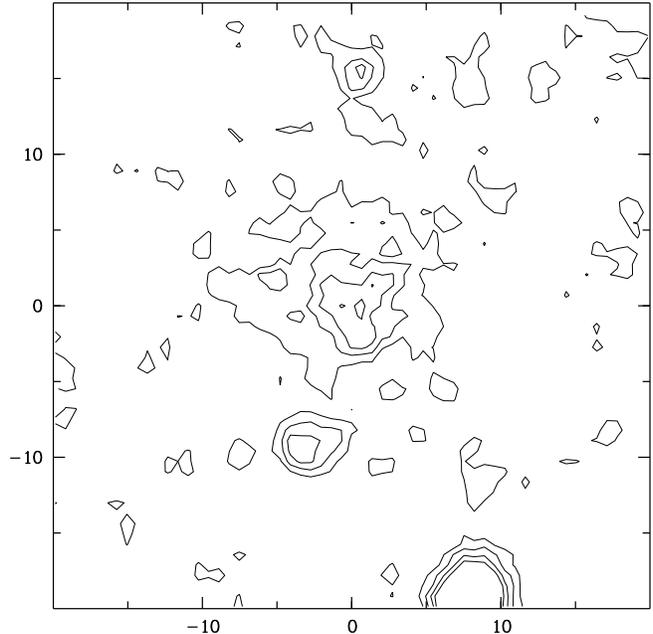} }

     \caption[]{ \object{AFGL~3099}. \emph{Left}: OHP 1.2~m image  in the V-band. 
     Field size: $117\arcsec \times 117\arcsec$. \emph{Right}: 
     Close-up isophotes of the V-band image. Field size:
     $40\arcsec \times 40\arcsec$. }
     \label{fig11}
     
     \end{figure*}

\begin{figure*}[!ht]
   
    \centerline{
     \includegraphics[width=0.48\linewidth]{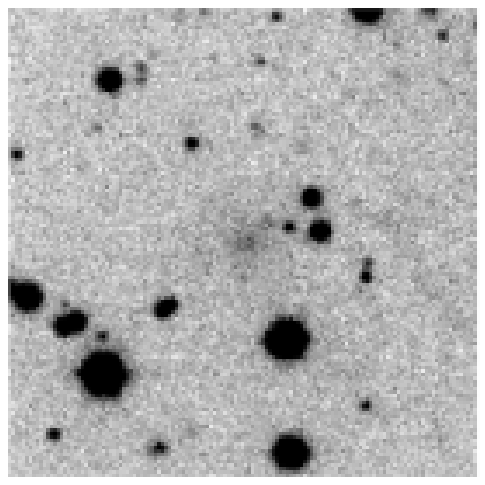}\hspace{0.2cm}
     \includegraphics[origin=c,angle=-90,width=0.48\linewidth]{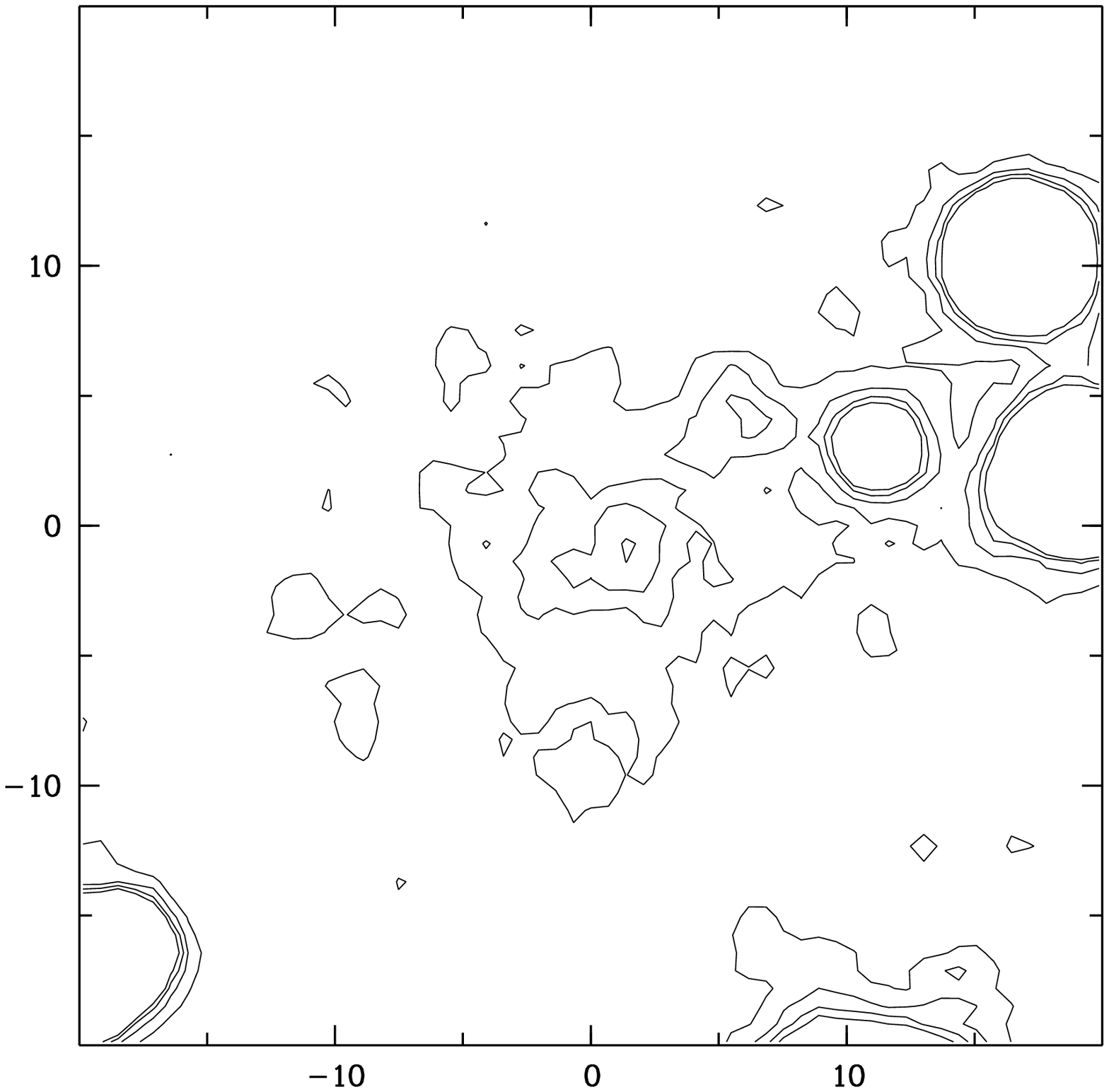}  }
     
     \caption[]{ \object{AFGL~3116}. \emph{Left}: OHP 1.20~m image in the B-band.   
     Field size: $117\arcsec \times 117\arcsec$. \emph{Right}:
     Close-up isophotes of the B-band image. Field size: $40\arcsec
     \times 40\arcsec$. }
     \label{fig12}
     
     \end{figure*}

\object{AFGL~3099} (Fig.\,\ref{fig11}) and \object{AFGL~3116}
 (Fig.\,\ref{fig12}) are the faintest envelopes
detected in the survey. They lie in relatively open star fields and in
azimuthally averaged radial profiles they can both be detected out to
$\sim 20\arcsec$ from the center. The isophotal contours suggest that
they may have interesting structure, but the signal-to-noise is too
low to make any firm statement on the geometry.

\section{Discussion}

\subsection{Envelope detection}

The stars for which extended envelopes are detected in scattered light
are generally those with the higher mass loss rates. The average mass
loss rates for the detections and non-detections are $1.6\pm0.5$ and
$0.7\pm0.2 \times 10^{-5}$~$M_{\odot}$\,yr$^{-1}$, respectively, with
only a few overlapping cases.  The dependence on mass loss rate is to
be expected on the basis of the following considerations.

For envelopes with relatively high mass loss rates in which the
stellar core is not seen, the intensity of the scattered light profile
increases toward the center, but reaches a plateau level at a radial
offset where the line of sight optical depth through the envelope reaches
$\sim 1$. Within the plateau, the intensity is roughly constant, or it
may dip at the center if the optical depth is large. The width (FWHM)
of the profile can be estimated by scaling the well-studied case of
\object{IRC+10216} (Mauron \& Huggins \cite{mh00}), assuming similar 
grain properties and dust-to-gas ratios:
\begin{displaymath} \Delta\theta = 5\arcsec\,
( \frac{\dot{M}}{10^{-5}\,\mathrm{M_\odot\ yr^{-1}} } ) (
\frac{V}{10\,\mathrm{km\ s^{-1}}} )^{-1} ( \frac{d}{1000\,\mathrm{pc}}
)^{-1}. \end{displaymath} The effect of increasing distance is to
reduce the angular size of the scattered light profile, but the
surface brightness of the plateau level is independent of distance for
a uniform galactic radiation field (ignoring interstellar
extinction). If this level is above the threshold and the core region
is resolved or partly resolved, the envelopes can be detected out to
quite large distances $\ga 1$~kpc.

For the envelopes detected here, the widths given by the above
equation (2--14\arcsec) are roughly comparable to the widths measured
in the profiles (4--12\arcsec), although there is no clear, overall
correlation, suggesting that individual variations dominate the small
dynamic range of the sample. Similarly, there is a variation in the
peak brightness by a factor of $\sim 10$. Besides intrinsic variation
among the envelopes (e.g., their internal structure, or scattering
properties), variation of the ambient galactic radiation field will
also affect the visibility.

In the case of \object{AFGL~3068}, the morphology gives clear evidence that
part of the illumination is contributed from a preferred
direction. The source could well be the bright star seen in 
Fig.\,\ref{fig07}.
 From APM photometry, the magnitude of this star in V is about
$+$12.0. If it is located at the same distance from us as \object{AFGL~3068}
(see Appendix A), the angular separation of 28\arcsec\ between the two
objects implies a physical separation of 0.15~pc. At this distance,
the V-band illumination of the nebula by the bright star would be a
factor of $\sim2$ larger than the average interstellar radiation
field.  No other stars around \object{AFGL~3068} appear to be good candidates
for the source of the additional illumination.

There are five objects whose envelopes were not detected in scattered
light. For one of these, \object{YY~Tri}, the star is not seen in the B or
V-bands. It probably has a thick envelope, but its large distance
yields an estimated width of $\sim {0}\farcs{5}$, which would make it
undetectable with the current observations. In the other four cases,
the estimated widths are somewhat larger (1--5\arcsec), but in each
the central star is very bright (16--18 magnitudes) relative the
surface brightness of scattered galactic light (e.g.,
25.2~mag~arcsec$^{-2}$ for \object{IRC+10216}). In these cases, the starlight
scattered by the envelope (see equation 1 of Mauron \& Le Borgne 
\cite{mauron86}),
which has a steep radial dependence, is lost in the extended stellar
profile of the ground-based observations. The limits on the galactic
contribution (at offsets of 5\arcsec\ or 10\arcsec) are therefore not
very sensitive compared to the cases where the starlight is suppressed
by a thick circumstellar envelope.

\subsection{Envelope time scales}

Table~4 summarizes our observations of the envelopes in terms of
their morphology, radial extent ($\theta_\mathrm{env}$), and the
corresponding expansion time scale ($\tau_\mathrm{env} =
d\,\theta_\mathrm{env}/V_\mathrm{exp}$), using the distances and
expansion velocities given in Table~1. We also give the morphology of
the core (and the corresponding time scale) in cases where this
differs from the extended envelope.

Table~4 includes data on the envelope of \object{IRC+10216} which was observed
by us in scattered light using ground-based and HST observations
(Mauron \& Huggins \cite{mh99}, \cite{mh00}). The extended envelope is detected out
to a distance of $\sim 200\arcsec$, similar to that seen in CO
(Huggins et al.\ \cite{huggins88}), and is roughly circular.  In the
central regions the envelope is bipolar. The WFPC2 F606W image shows a
classic two-lobed structure, separated by a dark lane. We characterize
the radial extent of the bipolar core by the radial distance to the outer 10\%
contour of each lobe ($1\farcs2$). Fainter light can be traced out to
$\sim 6\arcsec$. Similar dimensions are found with the F814W filter
(Skinner et al.\ \cite{skinner98}).

Table~4 also includes data on \object{CIT~6}. The extended envelope was not
detected in the ground-based observations reported in
Table~2. However, the stellar core was detected in HST WFPC2 images
reported by Trammell (\cite{trammell99}), Monnier et al. (\cite{monnier00}),
 and Schmidt et al. (\cite{schmidt02}). The wide-V (F555W) 
 image shows a classic two-lobed
structure, like that in \object{IRC+10216}, and we characterize the radial
extent by the distance to the outer 10\% contour of each lobe
($0\farcs3$); fainter light is seen out to beyond 1\arcsec. The
dimensions and appearance of the F675W image are similar. The larger
scale structure of \object{CIT~6} is not clear. At $\sim 10\arcsec$,
millimeter CN observations show asymmetry, with a major/minor axis
ratio of 1.4 oriented at a position angle $-25\degr$ (Lindqvist et
al.\ \cite{lindqvist00}), but at larger distances ($\sim 25\arcsec$) 
the envelope seen in CO appears circular (Neri et al.\ \cite{neri98}).  
In scattered light at 1~$\mu$, Schmidt et al.\ (\cite{schmidt02}) 
identify arcs between
1\arcsec--4\arcsec\ which may be similar to those seen in 
\object{IRC+10216}. 

The observations summarized in Table~4 show that scattered light
imaging can provide information about the mass loss on both short and
relatively long time scales, up to $\sim 15,000$~yr.  A detailed
analysis of the time dependent rate of the mass loss is beyond the
scope of the present report, but we illustrate the constraints
provided by the observations for the case of \object{IRC+10011} where, except
for the core region, the envelope appears spherically symmetric.

In the right hand panel of Fig.\,\ref{fig02} we compare the observed profiles of
\object{IRC+10011} with the results of a simple, analytical model for the
scattered light. The model assumes uniform external illumination of an
envelope with conservative scattering and an arbitrary phase
function. Under these conditions the scattered light profile is given
by: $ I_s = I_o(1-\mathrm{e}^{-\tau})$, where $I_o$ is the incident
intensity and $\tau$ is the optical depth along the line of sight (see
Mattila \cite{mattila70}). For a circumstellar envelope with a constant mass loss
rate, the optical depth at offset $x$ is proportional to $x^{-1}
\arccos(x/R)$ where $R$ is the outer radius. Such a model gives a
reasonable description of the well-observed profile of IRC+10216, and
the fit to \object{IRC+10011} is shown in Fig.\,\ref{fig02}. There is a degeneracy in
fitting the external radius which affects the outer parts of the
profile. The case shown is for $R=50\arcsec$; if the envelope extended
to a much larger radius there would be an extended $x^{-1}$ tail in
the profile with a value at 50\arcsec\ of $\sim 1.2$~ADU in V and
$\sim 1.7$~ADU in B. These values are relatively small, but above the
variation in the background.

The overall fits to the intensity profiles of \object{IRC+10011} are seen to be
quite good, and can be used to constrain large variations in the mass
loss rate. For example, Kemper et al. (\cite{kemper03}) have proposed that the
mass loss rate of \object{IRC+10011} may increase by a factor of 20 in the
range of $\sim 3$--9\arcsec, based on modeling the intensities of CO
spectra. Such a large variation would produce significant effects in
the scattered light profile (in regions where $\tau \la 1$). The
observations show no clear deviations from constant mass loss rate, so this
proposal can probably be ruled out, although this needs to be checked
with the development of more complete scattering models.  There may be
inhomogeneities in the envelope on small size scales, perhaps
associated with the possible narrow ring structure noted in \S4.3.1,
which might account for the CO observations.

The profiles of the other sources are all qualitatively similar to
those of \object{IRC+10011}, although \object{OH\,348.2$-$19.7} may have a steeper
fall-off in the outer regions. In no case is there evidence for
sudden, large changes in the mass loss rate on time scales longer than
that of the episodic multiple arcs ($\la 1000$~yr).

\subsection{Envelope morphology and shaping}

There are five AGB envelopes in Table~4 for which the structure is
well determined by the scattered light observations on both large and
small size scales.  It is remarkable that only one of the five
(\object{OH\,348.2$-$19.7}) is approximately spherical throughout. Two
(\object{IRC+10216} and \object{IRC+10011}) show roughly spherical extended envelopes, and
asymmetry or bipolarity close to the star; one
(\object{AFGL~2514}) shows an extended, elliptical envelope; and one
(\object{AFGL~3068}) shows a spiral pattern.

This sample of AGB envelopes is smaller than the sample in the CO
atlas of Neri et al.\ (\cite{neri98}) (see \S1) but the observations provide a
sharper picture of the morphology of the envelopes.  The fact that
nearly all show deviations from spherical symmetry develops the
scenario already implied by the case of the archetype \object{IRC+10216}
(Mauron \& Huggins \cite{mh00}; Skinner et al. \cite{skinner98}) that envelope shaping
begins on the AGB, before the transition to the proto-PN phase.

In addition to the frequent occurrence of asymmetry shown by the
scattered light observations, a second striking feature is the
diversity in the morphology of the envelopes. These show distinct
characteristics that can be used to help identify and constrain
possible shaping mechanisms.

\subsubsection{Core bipolarity}

There are two cases of spherical envelopes with core bipolarity,
\object{IRC+10216} and \object{IRC+10011}, as well as CIT~6 whose core is bipolar.  The
bipolarity is due to evacuated regions along the polar axes, and these
objects are likely precursors of proto-PNe and PNe with jets, which
have a similar geometry. The jets are usually ascribed to launching
from the accretion disk of a binary companion that is fed by the mass
loss from the AGB star (e.g., Frank \& Blackman \cite{frank04}). According to
Garcia-Segura et al.\ (\cite{garciasegura05}, and references therein) jets may also be
launched by the AGB star, but it has been argued by Soker (\cite{soker05}, and
references therein) that a binary companion is still needed in this
situation in order to spin-up the primary.

In the cases reported here, the extent of the bipolarity is small, and
the time scale is short (Table~4), much less than that of the extended
envelope.  In proto-PNe and young PNe, the jets are more extended.  In
the AGB stars the jets may be intrinsically weak, or they may have
just recently turned on at full power, and may be trapped in the AGB
envelope.  A model for a trapped jet has been proposed for \object{IRC+10011}
by Vinkovi\'{c} et al.\ (\cite{vinkovic04}). Our observations show that (within the
limitations of our small sample) core bipolarity is relatively common
in the late AGB phase and has a very short time scale. It seems
unlikely that the three cases observed here have just turned on at
full power. It is more probable that the jets are weak and/or
intermittent on the AGB, and gain in power or stability as the stars
evolve into the proto-PNe phase.

\subsubsection{Elliptical envelopes} 

The extended elliptical envelope found in \object{AFGL~2514} probably has a
three dimensional geometry that approximates a flattened (oblate)
ellipsoid or disk, with the axis of symmetry seen in projection as the
minor axis on the sky.  The observed major/minor axis ratio of $\sim
1.4$ then implies that the diameter/thickness ratio of the ellipsoid
is $\ga 1.4$, with the minimum value if the system is seen edge-on,
and larger values if the symmetry axis is inclined to the plane of the
sky. The corresponding density contrast between the equator and the
pole (at the same radius) is $q \ga 2$.

It is possible that single stars undergo asymmetrical mass loss and
produce elliptical envelopes (e.g., Dorfi \& Hofner \cite{dorfi96}), but the
interactions of the components of a binary system lead more naturally
to strong, asymmetrical mass loss with a preferred plane. There are
several ways in which the envelope in a binary system can acquire a
flattened morphology. One example is passage through a common envelope
phase. This may lead to the efficient ejection of the AGB envelope,
and in the simulations by Sandquist et al.\ (\cite{sandquist98}) 
flattened systems
emerge with density contrasts of $q \sim 5$. The common envelope
ejection is, however, a very rapid process, and this is contrary to
our observed scattered light profiles of \object{AFGL~2514} (Fig.\,\ref{fig06}) 
which show a smooth, asymmetric mass loss over a long time scale. 
So this process can be ruled out.

A second example of a binary shaping mechanism is the gravitational
focusing of the AGB wind by the companion.  Simulations of this have
been reported by Gawryszczak et al.\ (\cite{gaw02}) and Mastrodemos \& Morris
(\cite{mastrodemos99}).  For a given primary mass and wind velocity, the focusing
depends on the mass and separation of the secondary.  As a specific
example, model~11 of Mastredemos \& Morris, which has a primary of
mass 1.5~$M_\odot$ and a secondary of mass 1.0~$M_\odot$ with a
separation of 12.6~AU, produces a density contrast $q = 2$, close to
the minimum observed for \object{AFGL~2514}. A larger value of $q$, or a
similar value of $q$ with a smaller mass secondary, can be produced in
a closer binary system.

A possible alternative mechanism is the shaping of the AGB mass loss
by a stellar magnetic field. Matt et al.\ (\cite{matt00}) have shown that a
stellar dipole field can focus an initially isotropic wind toward the
equatorial plane, and they achieve a density contrast $q > 2$ for a
plasma $\beta$ parameter (the ratio of the gas to the magnetic
pressure) $\la 1$. For mass loss rates of
$10^{-5}$~$M_\odot$~yr$^{-1}$ this corresponds to field strengths at
the stellar surface of a few Gauss. Although this type of model is
plausible, Soker (\cite{soker05}) has pointed out that in cases where the
magnetic field plays a global role in the shaping, a binary companion
is necessary to maintain the field: so in this situation other binary
effects (such as those discussed above) also need to be taken into
account.

\begin{table}[!t]
\caption[]{Envelope morphology in  light}

	\begin{center}
        \begin{tabular}{llrrlr}
        \noalign{\smallskip}
        \hline
	\hline
        \noalign{\smallskip}

Star & Env.$^a$ & $\theta_\mathrm{env}$ & $\tau_\mathrm{env}$ & Core$^a$ & $\tau_\mathrm{c}$ \\
     &    &  $(\arcsec)$  & (yr)   & & (yr) \\       
\noalign{\smallskip}
\hline
\noalign{\smallskip}
\noalign{\smallskip}
\object{IRC+10216}       &  C s & 200 &  8700 & Bp     & 50 \\
\object{IRC+10011}       & C s? &  40 &  6800 & Bp     & 75 \\
\object{CIT 6}           & ? s? &  25$^b$& 3200$^b$ & Bp     & 40  \\
\object{AFGL 2155}       &  ?   &  18 &  5700 &$\ldots$  & \\
\object{OH\,348.2$-$19.7}& C    &  28 & 12900 & C      & \\
\object{AFGL 2514}       & E    &  40 &  7700 & E      & \\
\object{AFGL 3068}       & Sp   &  42 & 15200 & Sp     & \\
\object{AFGL 3099}       & ?    &  20 & 12200 &$\ldots$& \\
\object{AFGL 3116}       & ?    &  18 &  5600 &$\ldots$& \\
\noalign{\smallskip}
\noalign{\smallskip}
\hline
\end{tabular}
\end{center}
{\small $^a$ Morphology: C\,=\,circular, s\,=\,shells, E\,=\,elliptical, 
Sp\,=\,spiral, Bp\,=\,bipolar. $^b$ Based on CO.} \\

\end{table}

\subsubsection{Spiral envelopes} 

The spiral pattern in \object{AFGL~3068} (\S4.6) shows specific characteristics
that place tight constraints on a possible interpretation.
First, the spiral is one-armed.  Second, the pattern follows
the geometry of an Archimedes spiral, with an approximately constant
spacing between the arms. Third, the spiral is not planar, but is
composed of thin shells, as demonstrated by the intensity profiles.

These characteristics correspond closely to the spiral shocks found in
models of mass loss in binary systems investigated by Mastrodemos \&
Morris (\cite{mastrodemos99}). In this model, the mass-losing AGB star undergoes
reflex motion in the binary, which results in a three-dimensional
shock wave which propagates out through the envelope and creates the
spiral pattern. The pitch of the spiral directly reflects the period
of the binary and the shape is nearly spherical in the equatorial
plane so that the observed limb-brightened spiral is little affected
by moderate inclinations to the line of sight.

The radial pitch ($\lambda$) of the spiral that is given by the fit to
the data in \S4.6 is {2\farcs29}, which corresponds to $3.7\times
10^{16}$~cm at the distance of \object{AFGL~3068}.  For a velocity close to the
expansion velocity of the envelope ($V_{\mathrm exp}$ in Table~1),
which is likely to be the case far from the star where the structure
is frozen into the wind, the period is given directly by
$P=\lambda/V_{\mathrm exp} = 830$~yr.  Using Kepler's Law, the
separation of the system $a$ is given by $a=120
(\frac{M+m}{2.5})^{1/3} $~AU, where $M$ and $m$ are the masses (in
solar units) of the primary and secondary, respectively, and the
default is for $M=2$~$M_\odot$ and $m=0.5$~$M_\odot$.

This period and separation are larger than the spiral shock models
constructed by Mastrodemos \& Morris (\cite{mastrodemos99}), 
for which the maximum separation is $\sim 50$~AU, 
but they found no evidence that the effect
would not extend to wider systems. Overall, the evidence for 
interpreting the spiral as the interaction of a wide binary is
compelling.
 
\subsection{Binaries and PN formation} 

The results described above are of direct interest in connection with
the possible role of binaries in the formation of PNe, which is
currently an area of on-going debate 
(e.g., Meixner et al.\ \cite{meixner04}). One
line of evidence for the importance of binaries in PN formation is the
(possibly) large proportion of PN central stars with binary companions
(e.g., de Marco et al.\ \cite{demarco04}). A second line of evidence is the range
of morphologies of PNe which can be explained by the direct or
indirect results of binary interactions, and this has been explored
using population synthesis by Soker \& Rappaport (\cite{soker00}).

Given the nature of binary interactions, an important aspect of a
PN-binary scenario is that some of the effects of the interactions
should be present in the precursor phases. In this context, our
observations are in accord with a binary picture because most of the
well observed AGB envelopes reported here show asymmetries, which are
all consistent with direct or indirect binary interactions, as
discussed in the previous sections.

The diversity of the envelope morphologies reported here is also
strikingly in accord with the binary scenario because the interactions
depend in detail on the mass and separation of the companions.  We
expect that the different morphologies of the AGB envelopes reported
here will lead to different morphological types of fully formed PNe.
The mapping is probably not simple, and the size of our sample needs
to be significantly expanded before we can investigate the statistical
aspects of this in detail.

\section{Conclusions}

The observations reported here demonstrate that unique information on
the mass-loss of AGB stars can be obtained by imaging the
circumstellar envelopes in dust-scattered galactic light. The
envelopes can be detected out to large distances ($\sim 1$~kpc), and
the observations provide information on the mass loss history of the
stars over long time scales ($\sim 10,000$~yr).

The sample of AGB stars reported here is striking in revealing
asymmetries in four out of five cases where the envelope geometry can
be clearly determined. There is a diversity of morphology including a
flattened system, spiral structure, and asymmetry close to the
center. In all cases the asymmetries are consistent with the direct or
indirect effects of a binary companion.

Our results are in accord with a binary scenario for the shaping of
PNe, and further observations should lead to the possibility of
classifying in detail the early shaping of the envelopes of these PN
precursors.

\begin{acknowledgements}
 
  We acknowledge the use of the MIDAS software from ESO 
 which was used for the data processing. We thank an anonymous referee
 for the comments that helped improve the final version. 
 The HST data were obtained from
 the ESA/ESO ST-ECF Archive Center at Garching, Germany. 
 This work was supported in part by NSF grant AST 03-07277 (to PJH).

\end{acknowledgements}

\appendix
\section{\object{AFGL~3068}}

\begin{figure}[!ht]
     \resizebox{7.5cm}{!}{\rotatebox{0}{\includegraphics{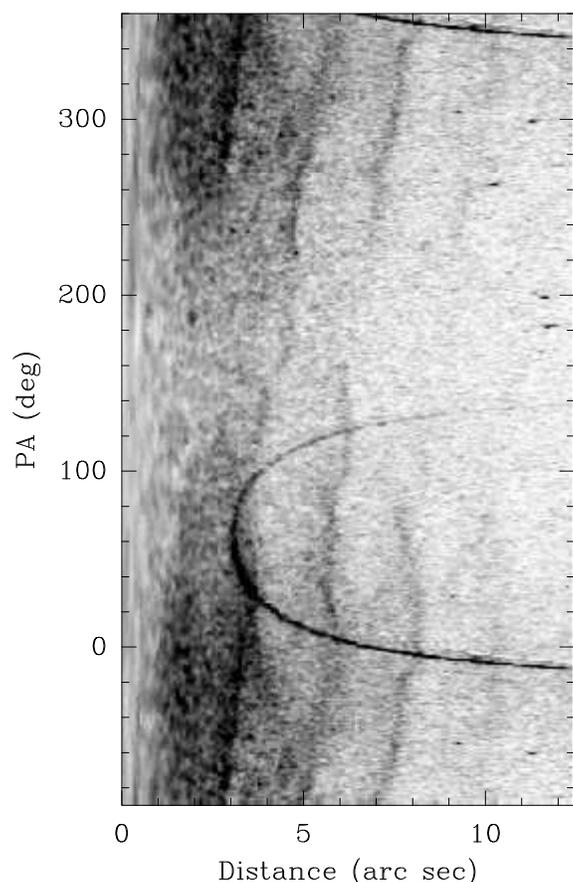}}}   
    \caption[]{Image of \object{AFGL~3068} of Fig.\,\ref{fig08} mapped into the position
      angle-radius plane. The sector $-$90\degr\ to 0\degr\ is
      repeated at 270\degr\ to 360\degr\ to show the continuity in
      position angle.  The curved line is the diffraction pattern of
      the bright star just outside the image.
      }
    \label{figa1}
   \end{figure}

 In order to demonstrate the spiral pattern in \object{AFGL~3068} we show in
Fig.\,\ref{figa1} the image of Fig.\,\ref{fig08}, mapped into the position angle--radius
plane. The sector from $-$90\degr\ to 0\degr\ is repeated in the
figure at 270\degr to 360\degr\ so that the continuity in position
angle can be traced. The center of the image is determined to an
accuracy of $\sim\pm{0}\farcs{15}$ from the position of the star seen
a longer wavelengths.

The signature of the single-armed spiral is the continuous tilted
feature in the image, which can be traced around 3--4 turns,
 exiting from the top section of the figure and continuing from
  the bottom section.
 There is substructure in the feature, part of which is contributed by the
asymmetrical illumination of the envelope.

If the bright star in Fig.\,\ref{fig07} illuminates the envelope, it can in
principle provide an independent distance to \object{AFGL~3068}. There is,
however, no parallax for the star, and unfortunately no spectroscopic
classification. Based on APM and 2MASS photometry, we find that the
best color matches to stars with known spectral classifications give a
spectral type of G5III or K1V. Using the absolute magnitudes given by
Cox et al.~(\cite{cox00}), the corresponding distance estimates to the star
are $1800\pm170$~pc for G5III, and $155\pm15$~pc for K1V. If the star
is a giant, it could illuminate the envelope, and the inferred
distance would be 66\% larger than the period-luminosity distance
given in Table~1. If the star is a dwarf, its illumination of the
envelope is ruled out. A spectroscopic classification would help
clarify the issue.



\Online
   \begin{figure*}[!ht]
    \resizebox{\hsize}{!}{
    {\includegraphics[origin=c,angle=-00,width=0.49\linewidth]{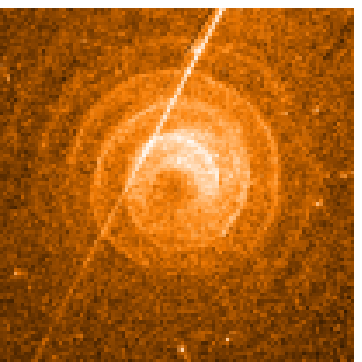}}\hspace{0.3cm} 
    {\includegraphics[origin=c,angle=-00,width=0.49\linewidth]{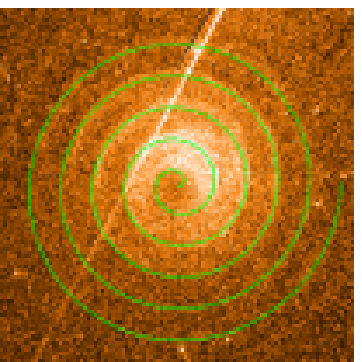}}}
     \caption[]{\object{AFGL 3068}. \emph{Left}:  Close-up of HST-ACS image
     in the F606W filter, lightly smoothed with a Gaussian of width
     (FWHM) 1.5 pixels. Field size: $25\farcs6 \times 25\farcs6$. 
     \emph{Right}: Same image, with Archimedes spiral fit. See text
     for details.}
     \label{figonline}
     \end{figure*}

\end{document}